\documentclass[aps,10pt,prd,twocolumn,superscriptaddress,preprintnumbers,%
showkeys,nofootinbib,showpacs]{revtex4-2}
\usepackage{bm}
\usepackage{graphicx}
\usepackage[hypertexnames=false, colorlinks=true, citecolor=blue, urlcolor=blue, linkcolor=black]{hyperref}
\usepackage{amsmath,amssymb,amsfonts,latexsym}
\begin{document}
\author{F. Gil-Dom\'inguez}
\email[ E-mail: ]{fernando.gil@ific.uv.es}
\affiliation{Instituto de F\'isica Corpuscular (IFIC) (centro mixto CSIC-UV), Institutos de Investigaci\'on de Paterna, C/Catedr\'atico Jos\'e Beltr\'an 2, 46980 Paterna, Valencia, Spain}
\author{J.~M.~Alarc\'on}
\email[ E-mail: ]{jmanuel.alarcon@uah.es}
\affiliation{Universidad de Alcal\'a, Grupo de F\'isica Nuclear y de Part\'iculas, Departamento de F\'isica y
Matem\'aticas, 28805 Alcal\'a de Henares (Madrid), Spain}
\author{C.~Weiss}
\email[ E-mail: ]{weiss@jlab.org}
\affiliation{Theory Center, Jefferson Lab, Newport News, VA 23606, USA}
\title{Proton charge radius extraction from muon scattering at MUSE \\
using dispersively improved chiral effective field theory}
\preprint{JLAB-THY-23-3834} 
\begin{abstract}
The MUSE experiment at Paul Scherrer Institute will perform the first measurement of low-energy
muon-proton elastic scattering (muon lab momenta 115--210 MeV) with the aim of determining the
proton charge radius. We study the prospects for the proton radius extraction using the theoretical
framework of Dispersively Improved Chiral Effective Field Theory (DI$\chi$EFT).
It connects the proton radii with the finite-$Q^2$ behavior of the form factors through complex
analyticity and enables the use of data up to $Q^2 \sim$ 0.1 GeV$^2$ for radius extraction.
We quantify the sensitivity of the $\mu p$ cross section to the proton charge radius,
the theoretical uncertainty of the cross section predictions, and the size of two-photon exchange
corrections. We find that the optimal kinematics for radius extraction at MUSE is at momenta
210 MeV and $Q^2 \sim$ 0.05--0.08 GeV$^2$. We compare the performance of electron and muon scattering
in the same kinematics. As a byproduct, we obtain explicit predictions for the $\mu p$
and $ep$ cross sections at MUSE as functions of the assumed value of the proton radius.
\end{abstract}
\maketitle
\section{Introduction}
The electromagnetic size is a fundamental characteristic of the proton observed
in nuclear and atomic physics.  It is quantified by the root-mean-squared radii
$r_E \equiv \sqrt{\langle r^2 \rangle_E}$ and $r_M \equiv \sqrt{\langle r^2
\rangle_M}$, defined by the derivatives of the electric and magnetic form
factors (FFs), $G_E$ and $G_M$, at momentum transfer $Q^2 = 0$,
see Ref.~\cite{Miller:2018ybm} for a review. The radii can be
determined experimentally either from elastic electron-proton ($ep$) or muon-proton
($\mu p$) scattering or from the nuclear corrections to the energy levels of
electronic or muonic hydrogen atoms.

The proton charge radius has been the object of extensive studies in the last
decade.  The extraction from muonic hydrogen measurements in 2010, $r_E =
0.84184(67)$~fm \cite{Pohl:2010zza}, differed by 5$\sigma$ from the CODATA value
accepted at the time, $r_E = 0.8768(69)$~fm \cite{Mohr:2008fa}, obtained from
electronic hydrogen and electron scattering data (``proton radius puzzle''). The
discrepancy motivated experimental and theoretical efforts aiming to
improve the extraction methods, quantify the uncertainties, and reconcile the
results; see Refs.~\cite{Pohl:2013yb,Carlson:2015jba} for reviews.
The questions raised include the performance of various methods for
extraction of the radius from scattering data, the comparison of scattering and
atomic results, and potential differences between electron and muon
interactions. The cumulative results from these studies tend to favor the
``smaller'' charge radius. However, one essential piece is still missing --
the extraction of the radius from low-energy $\mu p$ elastic scattering.

The MUSE experiment at Paul Scherrer Institute aims to perform the first precise
determination of the proton charge radius from $\mu p$ elastic scattering at muon
lab momenta 115--210 MeV \cite{MUSE:2017dod}. Good understanding of the
theoretical uncertainties is needed in order to optimize the extraction procedure
and assess the final error in the radius.  Important questions are the sensitivity
of the experimental observables to the proton radius, the theoretical uncertainty
in the relation between the proton radius and the finite-$Q^2$ FFs,
the size and uncertainty of two-photon exchange (TPE) corrections,
and the optimal muon energy and $Q^2$ range for constraining the radius.

The analytic properties of the proton FF play an essential role in the
radius extraction from scattering data. As a function of complex $Q^2$, the form
factor has singularities at $Q^2 < 0$, resulting from the $t$-channel exchange of
hadronic states (pions, resonances) between the electromagnetic current and the proton.
These singularities govern the behavior of the FF at $Q^2 > 0$, where it is
measured in scattering experiments. This structure implies a correlation between the
derivative of the FF at $Q^2 = 0$ and its values at finite $Q^2$,
which is essential for the radius extraction and must be implemented in the
theoretical analysis.

The recently developed method of Dispersively Improved Chiral Effective Field
Theory (DI$\chi$EFT) \cite{Alarcon:2017lhg,Alarcon:2018irp} combines dispersion
relations with dynamical input from chiral EFT to describe the nucleon FFs
at low $Q^2$ from first principles. It generates FFs
with correct analytic properties (position of singularities) and realistic
quantitative behavior (strength of singularities), which provide an excellent
description of scattering data up to $Q^2 \sim$ 1~GeV$^2$ \cite{Alarcon:2020kcz}.
It also quantifies the theoretical uncertainty of the FF calculations.
A special feature of this method is that it generates FF predictions that
depend on the assumed proton radius as a parameter. As such it explicitly realizes
the correlations between the proton radius and the finite-$Q^2$ behavior of the
FF. It permits the use of finite-$Q^2$ data for the radius extraction
with controlled uncertainties, which has many experimental and theoretical advantages.
The method has been used successfully for the extraction of the proton electric and
magntic radii from electron scattering data \cite{Alarcon:2018zbz,Alarcon:2020kcz}.

In this work we use DI$\chi$EFT to study the prospects for proton radius determination
in $\mu p$ elastic scattering at MUSE and optimize the extraction procedure.
We compute the $\mu p$ cross section with the DI$\chi$EFT FFs, quantify the theoretical
uncertainties and TPE corrections, and evaluate the sensitivity to the proton radius.
Specifically, we attempt to answer the following questions:
\begin{enumerate}
\item What is the theoretical sensitivity of the $\mu p$ cross section in MUSE kinematics
to the proton radius?
\item What are the theoretical uncertainties in the $\mu p$ cross section resulting
from the DI$\chi$EFT FF predictions and from TPE corrections?
\item What kinematic range in beam energy and $Q^2$ has the most impact on the radius extraction?
\item What are the differences between $ep$ and $\mu p$ scattering in
radius extraction in MUSE kinematics?
\end{enumerate}
We demonstrate that the radius extraction is characterized by a trade-off between several effects -- the
sensitivity of the cross section to the radius, the theoretical uncertainty in the FF
predictions for a given radius, and the size and kinematic dependence of TPE
corrections \cite{Alarcon:2018zbz,Alarcon:2020kcz}.
We determine the optimal kinematics for radius extraction at MUSE based on these considerations.
In addition, we provide predictions of the expected $\mu p$ and $ep$ cross sections
for the nominal value of the proton radius.

\section{Methods}

\subsection{Lepton-proton elastic scattering}
The elastic lepton-proton scattering process $l(k) + p(p) \rightarrow l(k') + p(p')$, where
$l = \mu^\mp$ or $e^\mp$, is described by the invariant variables
\begin{align}
s \equiv (k + p)^2,
\hspace{2em}
Q^2 = -t \equiv -(k - k')^2.
\end{align}
In the initial proton rest frame (lab frame), the initial and final muon momenta are
$\bm{k}$ and $\bm{k}'$, the energies are $\omega \equiv \sqrt{|\bm{k}|^2 + m^2}$
and $\omega' \equiv \sqrt{|\bm{k}'|^2 + m^2}$, and the invariants are given by
\begin{align}
s = M^2 + 2 M \omega + m^2,
\hspace{2em}
Q^2 = 2 M (\omega - \omega'),
\end{align}
where $m$ is the lepton mass and $M$ the proton mass. The scattering angle $\theta_{\rm lab} =
\textrm{angle}(\bm{k}', \bm{k})$ is related to the final lepton energy and momentum by
\begin{align}
\cos\theta_{\rm lab} = \frac{\omega\omega' - m^2 - M (\omega - \omega')}{|\bm{k}||\bm{k}'|}.
\end{align}
The kinematic range of the momentum transfer accessible at a given initial lepton momentum is
\begin{align}
0 \; \leq \; Q^2 \; \leq \; \frac{4 M^2 |\bm{k}|^2}{s} \equiv Q^2_{\rm max}.
\label{Q2_limits}
\end{align}

In the one-photon-exchange approximation, the differential cross section for unpolarized scattering
is given by (see e.g. Ref.~\cite{Tomalak:2018jak})
\begin{align}
\frac{d\sigma_{1\gamma}}{dQ^2} &= \frac{\pi \alpha^2}{2M^2 |\bm{k}|^2}
\frac{(\epsilon/\tau_P) \, G_E^2 + G_M^2}
{1-\epsilon_T} .
\label{Eq:X-sec_1gamma}
\end{align}
Here $\alpha$ is the fine structure constant, and $G_{E, M} \equiv G_{E, M}(Q^2)$
are the electric and magnetic Sachs FFs of the proton. $\epsilon$ is the
virtual photon polarization parameter and given by
\begin{align}
&\epsilon = \frac{\displaystyle Q^2_{\rm max} - Q^2 + \frac{m^2}{s} (4 M^2 + Q^2)}
{\displaystyle Q^2_{\rm max} - Q^2 + \frac{Q^2}{2s} (4 M^2 + Q^2)}, 
\end{align}
and $\tau_P \equiv Q^2/(4M^2)$. $\epsilon/\tau_P$ is the ratio of the
fluxes of longitudinal and transverse polarized photons in the one-photon-exchange approximation.
$\epsilon_T$ is the degree of linear polarization of the transverse photons,
\begin{align}
&\epsilon_T = \frac{\displaystyle Q^2_{\rm max} - Q^2}
{\displaystyle Q^2_{\rm max} - Q^2 + \frac{Q^2}{2s} (4 M^2 + Q^2)}, 
\end{align}
and is bounded by $0 \leq \epsilon_T < 1$. In the case of zero lepton mass (as usually assumed in
electron scattering) $\epsilon = \epsilon_T$, but for non-zero lepton mass (muon scattering) there
are important differences. $\epsilon$ attains values $> 1$ at $Q^2 = 0$, and remains nonzero
at $Q^2 = Q^2_{\rm max}$,
\begin{align}
\epsilon (Q^2 = 0) &= \frac{\omega^2}{|\bm{k}|^2} \; > \; 1,
\\
\epsilon (Q^2 = Q^2_{\rm max}) &= \frac{m^2 s}{2 M^2 |\bm{k}|^2} \; > \; 0.
\label{epsilon_Q2max}
\end{align}

Two-photon exchange (TPE) corrections play an important role in the analysis of low-energy lepton-proton
elastic scattering, see Refs.~\cite{Carlson:2007sp,Arrington:2011dn} for a review.
At order $\alpha^3$, the correction arises from the interference between the two-photon
and one-photon exchange amplitudes and is usually included through a multiplicative factor
modifying the one-photon exchange cross section,
\begin{align}
\frac{d\sigma}{dQ^2} \approx  \frac{d\sigma_{1\gamma}}{dQ^2} \, \left( 1 + \delta_{2\gamma} \right).
\label{Eq:TP-Correction}
\end{align}
The correction $\delta_{2\gamma}$ for $\mu p$ scattering has been computed in several theoretical approaches,
such as dispersion theory \cite{Tomalak:2015hva,Tomalak:2018jak} and chiral
effective field theory \cite{Talukdar:2019dko,Talukdar:2020aui,Cao:2021nhm}. In this work we use the
results of Ref.~\cite{Tomalak:2015hva}, which give corrections $\delta_{2\gamma} \lesssim 0.5\%$
in the kinematic range of the MUSE experiment. While in Ref.~\cite{Tomalak:2015hva} the inelastic
contribution to the dispersion integral for $\delta_{2\gamma}$ was computed in forward kinematics,
the analysis of Ref.~\cite{Tomalak:2018jak} showed that this approximation is accurate within $10\%$. 
\subsection{DI$\chi$EFT representation of form factors}
%
%
\begin{figure*}[t]
\includegraphics[width=.7\textwidth]{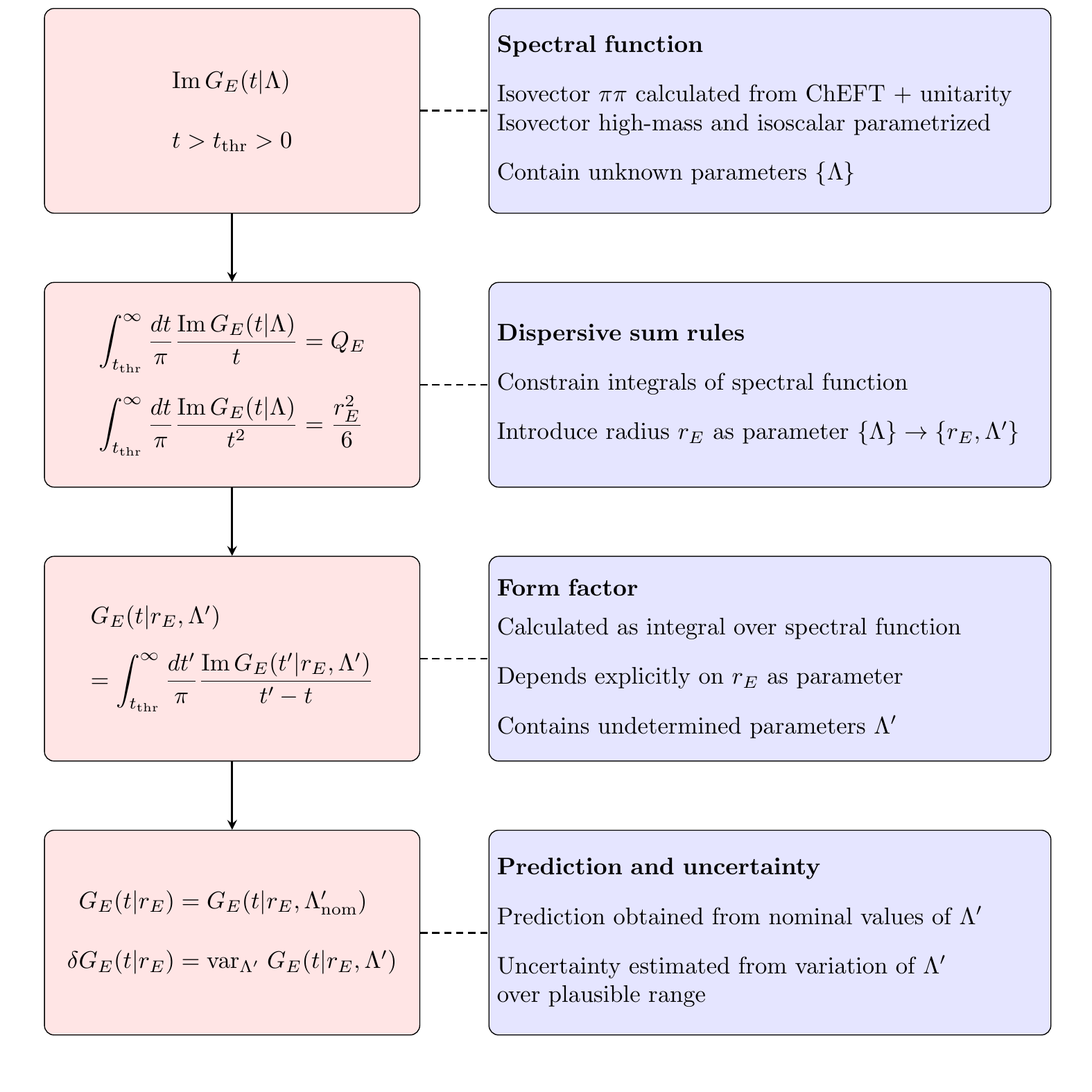} 
\caption{Flowchart of the DI$\chi$EFT description of the nucleon FFs. Shown is the case of the proton
electric FF $G_E$; the same flow applies to the neutron electric FF. In the case of the magnetic FF $G_M$,
the right-hand side of the dispersive sum rules is given by the magnetic moment $\mu$ and the
magnetic radius $r^2_M$.}
\label{Fig:dicheft_schematic}
\end{figure*}

The foundations of the DI$\chi$EFT method and its applications are described in detail in
Refs.~\cite{Alarcon:2017ivh,Alarcon:2017lhg,Alarcon:2018irp}. Here we provide a brief summary,
emphasizing the information flow (parameters) and other features relevant to proton radius extraction.

DI$\chi$EFT is based on dispersion theory, in which the FFs $G_{E, M}(t)$ at spacelike momentum transfer
$t < 0$ are represented as integrals over their imaginary parts $\textrm{Im}\, G_{E, M}(t)$
on the cut at timelike $t > 0$, the so-called spectral functions. The main steps in the DI$\chi$EFT
construction of the spectral functions and the FFs are summarized in Fig.~\ref{Fig:dicheft_schematic}.
In the following we describe the steps for $G_E$; the ones for $G_M$ are similar.

In the first step, one constructs the spectral function. The proton FF has an isovector and
isoscalar component, $G_E \equiv G_E^V +G_E^S$. The isovector FF $G_E^V$ has the two-pion cut
at $t > t_{\rm thr} = 4 M_\pi^2$. The isovector spectral function is represented as the sum of two parts,
\begin{align}
\text{Im}\, G_E^V(t) &= \text{Im}\, G_E^V(t)[\pi\pi] + \text{Im}\, G_E^V(t)[\textrm{high-mass}].
\label{spectral_pipi_highmass}
\end{align}
The $\pi\pi$ part covers the region $4 M_\pi^2 \leq t < t_{\rm max} \approx 1$ GeV$^2$ and
is computed theoretically, using the elastic unitarity relation in the $\pi\pi$ channel and relativistic
chiral EFT for the $\pi N$ amplitudes. This approach includes the $\rho$ resonance in the $\pi\pi$ channel
and delivers realistic $\pi\pi$ spectral functions, which is essential for proton radius extraction (see below).
The free parameters entering in this part are the low-energy constants of the chiral EFT calculation;
in the present partial N$^2$LO implementation this is one parameter, $\lambda$ \cite{Alarcon:2018irp}.
The high-mass part of the spectral function covers the region $t > 1$ GeV$^2$ and is parametrized by a single
effective pole $\pi a_1 \delta (t - t_1)$; this form has been shown to be sufficient for the dispersive
analysis of low-$t$ FFs, which see only the overall spectral strength
in the high-mass region \cite{Alarcon:2018irp}. The free parameters entering in this part
are the pole strength $a_1$ and pole position $t_1$. The isoscalar FF $G_E^S$ has a three-pion cut,
and the spectral function is represented as
\begin{align}
\text{Im}\, G_E^S(t) &= \text{Im}\, G_E^S(t)[\pi\pi\pi] + \text{Im}\, G_E^S(t)[\textrm{high-mass}]. \hspace{-.5em}
\label{spectral_3pi_highmass}
\end{align}
The $\pi\pi\pi$ part is overwhelmingly concentrated in the $\omega$ resonance
and parametrized by a pole $\pi a_\omega \delta (t - M_\omega^2)$. The high-mass part is
parametrized by an effective pole, whose position can be taken as the $\phi$ mass,
$\pi a_\phi \delta (t - M_\phi^2)$. The free parameters entering in the isoscalar spectral function
are the pole strengths $a_\omega$ and $a_\phi$. Altogether, this step results in a theoretical
parametrization of the spectral function
\begin{align}
\text{Im}\, G_E (t|\Lambda) = \text{Im}\, G_E^V(t) + \text{Im}\, G_E^S(t),
\end{align}
where $\{ \Lambda \}$ collectively denotes the free parameters; in the present analysis
$\{ \Lambda \} = \{ \lambda, a_1, t_1; a_\omega, a_\phi\}$.

In the second step, one imposes the dispersive sum rules for the proton charge and radius
\begin{align}
&\frac{1}{\pi}\int_{t_{\rm thr}}^\infty dt \frac{\text{Im}\, G_E(t|\Lambda)}{t} = Q_E,
\label{sumrule_charge}
\\
&\frac{1}{\pi}\int_{t_{\rm thr}}^\infty dt \frac{\text{Im}\, G_E(t|\Lambda)}{t^2} = \frac{r^2_E}{6},
\label{sumrule_radius}
\end{align}
where $Q_E = 1$ is the proton charge and $r_E^2 > 0$ the proton charge radius squared. (The same relations
are imposed for the neutron electric FF, in which case $Q_E = 0$ and $r_E^2 < 0$ is the negative neutron charge
radius squared.) These relations express the FF at $t = 0$ and its derivative as integrals over the spectral function.
One uses them to constrain the parameters in the spectral function. In particular, Eq.~(\ref{sumrule_radius})
is valid for any assumed value of the proton charge radius $r_E$, and one can use it to express one of
(or a combination of) the original parameters in terms of the radius, i.e., to introduce the
radius as a parameter:
\begin{align}
\{ \Lambda \} \rightarrow  \{ r_E, \Lambda' \} .
\end{align}
In this way one obtains a set of spectral functions that depend explicitly on the assumed radius, as
well as on the remaining parameters $\Lambda'$
\begin{align}
\text{Im}\, G_E (t | r_E, \Lambda').
\end{align}
In the present analysis we use Eqs.~(\ref{sumrule_charge}) and (\ref{sumrule_radius}) for the
proton spectral function (and the same relations for the neutron) to fix the chiral low-energy
constant $\lambda$ and the effective pole strengths $a_1, a_\omega, a_\phi$, retaining the
isovector effective pole position $t_1$ as the only undetermined parameter.

In the third step, one computes the spacelike FFs ($t < 0$, or $Q^2 > 0$) as the dispersion
integral with the spectral function,
\begin{align}
G_E (t| r_E, \Lambda')
= \frac{1}{\pi}\int_{t_{\rm thr}}^\infty dt' \frac{\text{Im} \, G_E (t'| r_E, \Lambda')}{t' - t}.
\label{dispersion_ff}
\end{align}
The FF thus obtained depends on the assumed radius $r_E$ and the undetermined parameters $\Lambda'$.
The nominal prediction for the FF with assumed radius $r_E$ is obtained with the nominal values of
$\Lambda'$,
\begin{align}
G_E (t| r_E) = G_E (t| r_E, \Lambda'_{\rm nom}) .
\label{FF_nominal}
\end{align}
In the last step, the theoretical uncertainty of the FF with assumed radius
$r_E$ is estimated by varying $\Lambda'$ over a plausible range,
\begin{align}
\delta G_E (t| r_E) = \textrm{var}_{\Lambda'} \; G_E (t| r_E, \Lambda') .
\label{FF_uncertainty}
\end{align}
In this way one obtains a nominal prediction and a theoretical uncertainty estimate for the form
factor with any given assumed radius. In the present analysis, the undetermined parameter is the
position of the isovector high-mass pole; its nominal value is $t_1 =$ 2.1 GeV$^2$, and the plausible range of
variation for the uncertainty estimate is $t_1 =$ 1.4--2.8 GeV$^2$ \cite{Alarcon:2018zbz}.

The magnetic FF $G_M^V$ and its uncertainty are constructed by an analogous procedure.
The dispersive sum rules for $G_M$ now involve the magnetic moment $\mu$ and the magnetic radius
$r_M^2$ see Ref.~\cite{Alarcon:2018irp} for details. A computer code generating the radius-dependent
DI$\chi$EFT FFs $G_{E, M}$ used in the present analysis is available in the
supplemental materials of Ref.~\cite{Alarcon:2020kcz}. 

%
%
\subsection{Proton radius extraction}
\begin{figure}[t]
\includegraphics[width=1.0\linewidth]{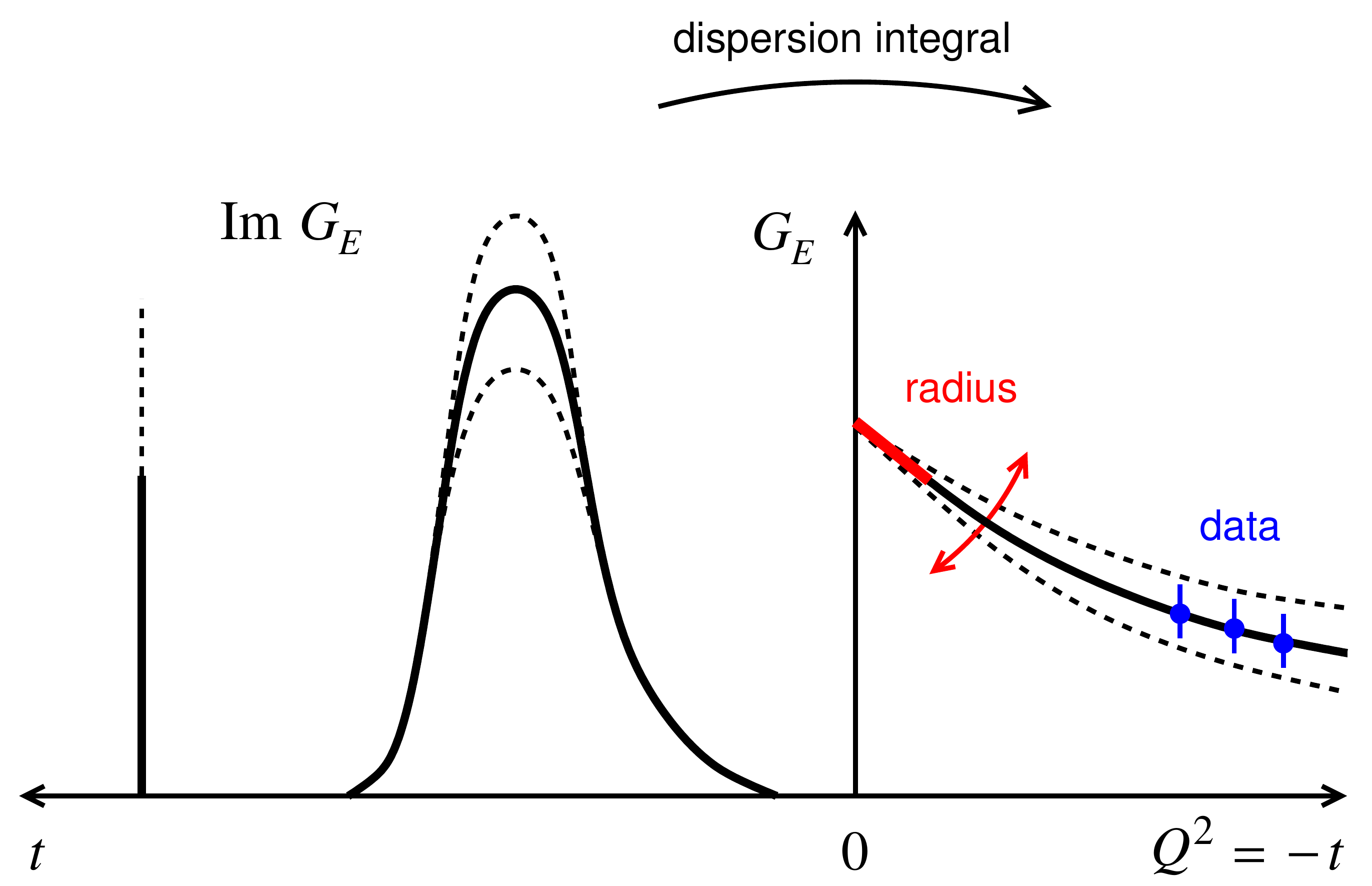} 
\caption{Illustration of the correlation between the proton radius and the spacelike FF, resulting from
analyticity and the information flow in DI$\chi$EFT.
The assumed value of the radius constrains the spectral function through the sum rule Eq.~(\ref{sumrule_radius}).
The corresponding FF at $Q^2 > 0$ is produced by the dispersion integral Eq.~(\ref{dispersion_ff}).
Variation of the radius causes variation of the spectral function and the corresponding FF.
(The graph shows only the isovector part of the spectral function at $t > 0$.)}
\label{Fig:correlation}
\end{figure}
The DI$\chi$EFT representation of the FFs enables a new theory-guided method of proton radius extraction
(see Fig.~\ref{Fig:correlation}) \cite{Alarcon:2018zbz,Alarcon:2020kcz}. 
For each assumed value of the proton radius, the theory generates a spectral function whose features
(height of the $\rho$ resonance peak, strength of effective poles) quantitatively depend on the value of the radius.
The dispersion integral projects these features into the spacelike region, up to spacelike momentum
transfers of the order $Q^2 \sim M_\rho^2$ and beyond. This effectively correlates the assumed value
of the radius with the behavior of the spacelike FF at finite $Q^2$ of this order.
The correlation described here is based on complex analyticity and the particular information flow in the
DI$\chi$EFT calculation and extends far beyond what one could infer from the series expansion in $Q^2$
with a given first derivative. Altogether, this allows one to recruit FF data of the order
$Q^2 \sim M_\rho^2$ and beyond for constraining the proton radii.

Radius extraction using DI$\chi$EFT proceeds as follows.
For a range of assumed radii, one generates the DI$\chi$EFT FFs as functions of $Q^2$, including their
theoretical uncertainties resulting from the undetermined parameters, Eqs.~(\ref{FF_nominal}) and
(\ref{FF_uncertainty}). From these FFs one predicts the cross section for the given assumed radius,
including its theoretical uncertainty from the FFs and two-photon exchange corrections.
For a given experimental setup (kinematic coverage, statistical and systematic errors)
one can then assess how data in a given range of energies and $Q^2$ can constrain the radii.
The optimal range is determined by a trade-off between the sensitivity of the DI$\chi$EFT FFs
to the value of the radius, the theoretical uncertainty of the DI$\chi$EFT FFs, the two-photon
exchange effects, and the precision of the data \cite{Alarcon:2018zbz,Alarcon:2020kcz}. 
The actual radius can then be determined by a fit in this optimal range, taking into account
all the uncertainties. In the following we apply this method to $\mu p$ scattering at MUSE and
discuss the prospects for proton radius extraction.

The DI$\chi$EFT method offers several advantages compared to other methods of proton radius extraction.
Compared to empirical fits (polynomials, splines), the DI$\chi$EFT method incorporates the analytic
structure of the FFs, which includes both the position of the singularities at $t > 0$ and the quantitative
distribution of strength in the spectral function. The analytic structure governs the global behavior
of the FF, which is difficult to implement in approaches based on polynomial expansions because of strong
correlations between higher-order coefficients (analyticity effectively controls the ``collective
behavior'' of higher derivatives of the FF at $Q^2 = 0$ \cite{Alarcon:2017lhg}).
Compared to traditional dispersion analysis \cite{Hohler:1976ax,Belushkin:2006qa,Lorenz:2012tm,Hoferichter:2016duk},
the DI$\chi$EFT method allows the strength of the
isovector spectral function in the $\rho$ resonance region to vary with the proton radius in a theoretically
controlled manner, providing critical flexibility for fitting the spacelike FF data and recruiting
them for radius determination. In traditional dispersive fits the $\pi\pi$ part of the isovector
spectral function is completely fixed by theory, and the spacelike FF data only constrain the high-mass part
of the spectral function, which restricts the interplay of the FF data with the proton radius 
(the $\rho$ resonance region of the spectral function accounts for about half the value
of the proton radius in Eq.~(\ref{sumrule_radius}) \cite{Alarcon:2017lhg}).
\section{Analysis}

\subsection{Sensitivity of $\mu p$ cross section to proton radius}
\label{subsec:sensitivity}
We now apply the DI$\chi$EFT framework to study the prospects for proton radius extraction at MUSE.
In the first step, we study the sensitivity of the $\mu p$ elastic scattering cross section to the
proton electric radius and compare it with the theoretical uncertainties resulting from the DI$\chi$EFT
FF predictions, from two-photon exchange corrections, and from the magnetic FF contributions.

\begin{figure}[t]
\includegraphics[width=0.9\columnwidth]{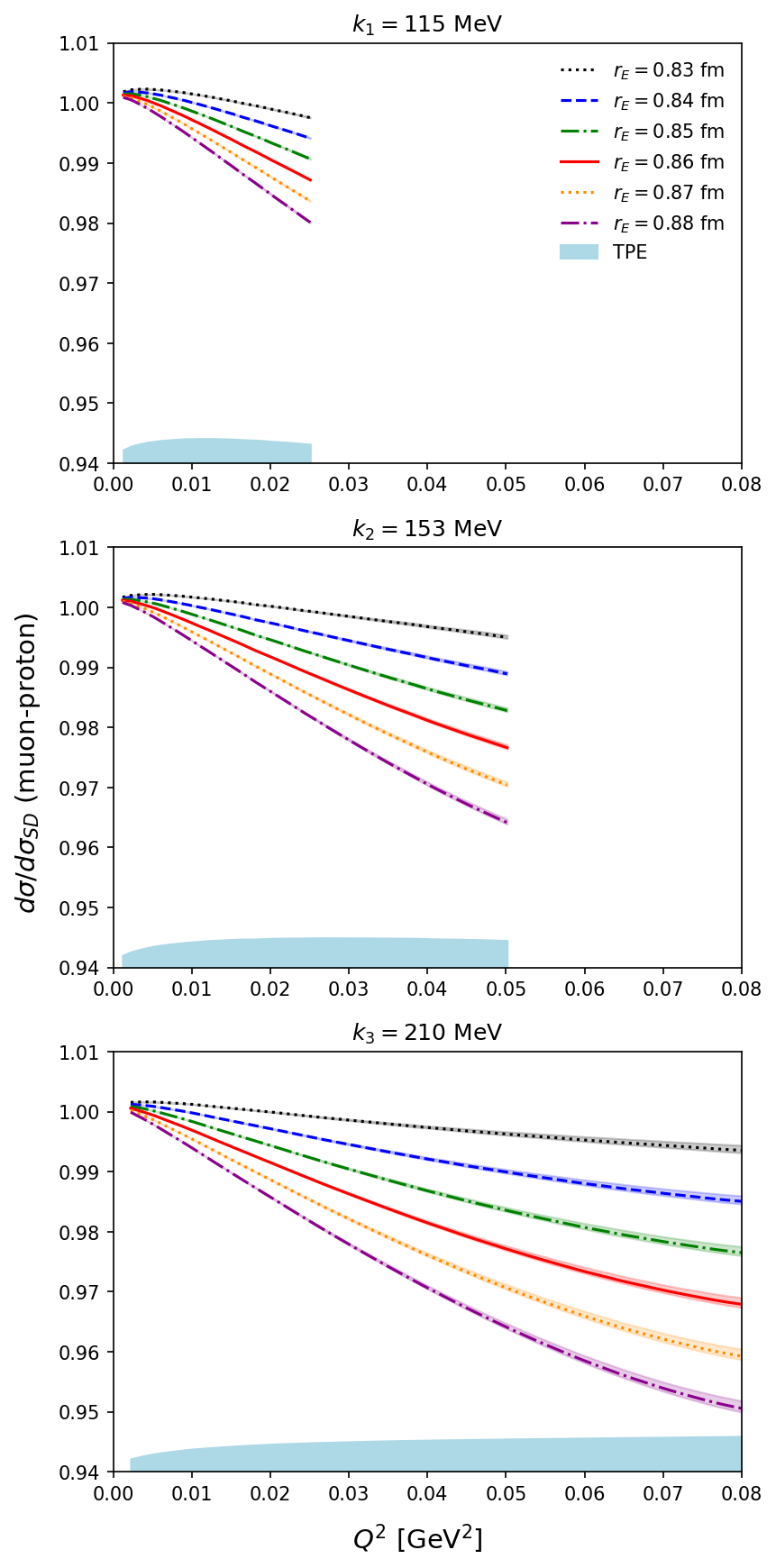} 
\caption{DI$\chi$EFT predictions for the $\mu^- p$ differential cross section at MUSE for several assumed
values of the proton radius. The cross section predictions include the TPE correction,
Eq.~(\ref{Eq:TP-Correction}), and are normalized to the standard dipole cross section without TPE effects.
{\it Lines (solid, dashed, dotted, dahed-dotted):}
Nominal DI$\chi$EFT predictions for the assumed value of the proton radius (see legend).
{\it Shaded bands around lines:} Intrinsic theoretical uncertainty of DI$\chi$EFT predictions,
unrelated to assumed proton radius.
{\it Shaded band at bottom:} TPE contribution to cross section \cite{Tomalak:2015hva}.}
\label{Fig:cross_section_radii_mup}
\end{figure}

To exhibit the various effects, we generate a set of DI$\chi$EFT FF predictions by varying the proton electric
radius over the range $r_E =$ 0.83--0.88 fm in steps $\Delta r_E =$ 0.01 fm, and evaluate the $\mu p$ elastic
scattering cross section with each of these FFs (the magnetic radius is kept at its nominal value; the role
of the magnetic FF in the cross section is discussed below). We include in the cross section the TPE
correction of Ref.~\cite{Tomalak:2015hva}. Figure~\ref{Fig:cross_section_radii_mup} shows the 
predicted cross sections for various incident muon momenta $k \equiv |\bm{k}|$,
as functions of $Q^2$. The lines show the
cross section obtained with the nominal DI$\chi$EFT FF predictions for each value of the radius,
Eq.~(\ref{FF_nominal}); the associated bands show the variation due to the theoretical uncertainty
of the DI$\chi$EFT FF predictions for the given radius, Eq.~(\ref{FF_uncertainty}). The bands at the bottom
of the plots show the absolute size of the TPE correction in the cross section predictions
(note that this is the overall size of the TPE correction, not its theoretical uncertainty).
The standard dipole cross section ($\sigma_{SD}$) used for normalization is the one-photon exchange
cross section evaluated assuming the standard dipole $Q^2$-dependence
$\propto (1 + Q^2/0.71 \, \textrm{GeV}^2)^{-2}$ for both FFs $G_{E, M}$.
%
%

One observes: (a)~The sensitivity of the cross section to the proton radius increases with $Q^2$ and
with the beam momentum $k$, because the separation of the FF predictions with different radii
increases with $Q^2$ \cite{Alarcon:2018zbz,Alarcon:2020kcz}. At the highest beam momentum,
$k =$ 210 MeV, the relative variation of the cross section reaches $\Delta\sigma/\sigma \sim$ 1\% for
$Q^2$ at the upper end of the range shown here.
(b)~The theoretical uncertainty of the cross section predictions for given radius also increases with $Q^2$
\cite{Alarcon:2018zbz,Alarcon:2020kcz}. Overall, the theoretical uncertainty is substantially smaller
than the relative variation of the cross section for $\Delta r_E$ = 0.01 fm over the kinematic range
shown here. (c)~The magnitude of the TPE correction does not vary strongly with $Q^2$ and $k$ over the
range covered here. At the upper end of the $Q^2$ range, the magnitude of the TPE correction is
comparable to the relative variation of the cross section with $\Delta r_E$ = 0.01 fm.
This clearly shows the importance of the TPE correction for radius extraction.

%
%
\begin{figure}[t]
\includegraphics[width=0.8\linewidth]{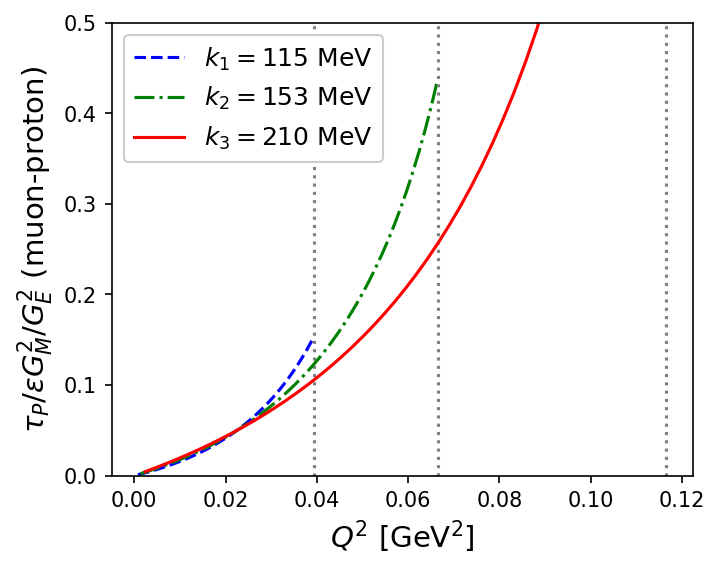} 
\caption{Ratio of magnetic and electric contributions to the $\mu p$ elastic scattering cross section,
$(\tau_P / \epsilon) G_M^2/G_E^2$, in MUSE kinematics.
The vertical dotted lines represent the kinematic upper limits of $Q^2$ at the
given beam momentum $k$, Eq.~(\ref{Q2_limits}).}
\label{Fig:GM_GE_ratio_mup}
\end{figure}
We also need to consider the uncertainties resulting from the contribution of the magnetic FF
to the $\mu p$ elastic scattering cross section. This is particularly important, as with the
DI$\chi$EFT framework we can recruit data at higher $Q^2$ for radius extraction, comparable to
$Q^2_{\rm max}$ at the given $k$. Figure~\ref{Fig:GM_GE_ratio_mup} shows the ratio of magnetic
and electric contributions to the one-photon exchange cross section, $(\tau_P / \epsilon) G_M^2/G_E^2$,
in MUSE kinematics. One sees that the ratio depends mainly on $Q^2$, having values $\sim$0.1
at $Q^2$ = 0.04 GeV$^2$ and reaching $\sim$0.4 at $Q^2$ = 0.08 GeV$^2$. Overall, the magnetic
contributions to the cross section are limited in all kinematic settings. Using the DI$\chi$EFT
framework and the results of the analysis of $ep$ scattering data of
Ref.~\cite{Alarcon:2020kcz}, we have computed the effect of the experimental uncertainties
of $G_M$ on the $\mu p$ cross section predictions in MUSE kinematics. We observe a maximum
variation in the cross sections of the order of $0.04\%$ at the highest $Q^2$, which is small
compared to the variation of $\sim$1\% resulting from a change of the electric radius
$\Delta r_E/r_E$ = 1\%. We conclude then that the current experimental uncertainties in $G_M$
do not limit the extraction of the proton electric radius from the $\mu p$ scattering data at
the accuracy considered here.
\subsection{Optimal kinematics for proton radius extraction}
In the second step, we discuss the optimal kinematic range for the radius extraction at MUSE.
It is determined by the trade-off between the sensitivity of the cross section to the radius,
the theoretical uncertainties of the DI$\chi$EFT FF predictions and the TPE corrections,
and the experimental errors of the cross section measurement. While the experimental
errors can only be estimated at present, some interesting conclusions can already be
obtained at this stage.

%
%
\begin{figure}[t]
\includegraphics[width=0.95\columnwidth]{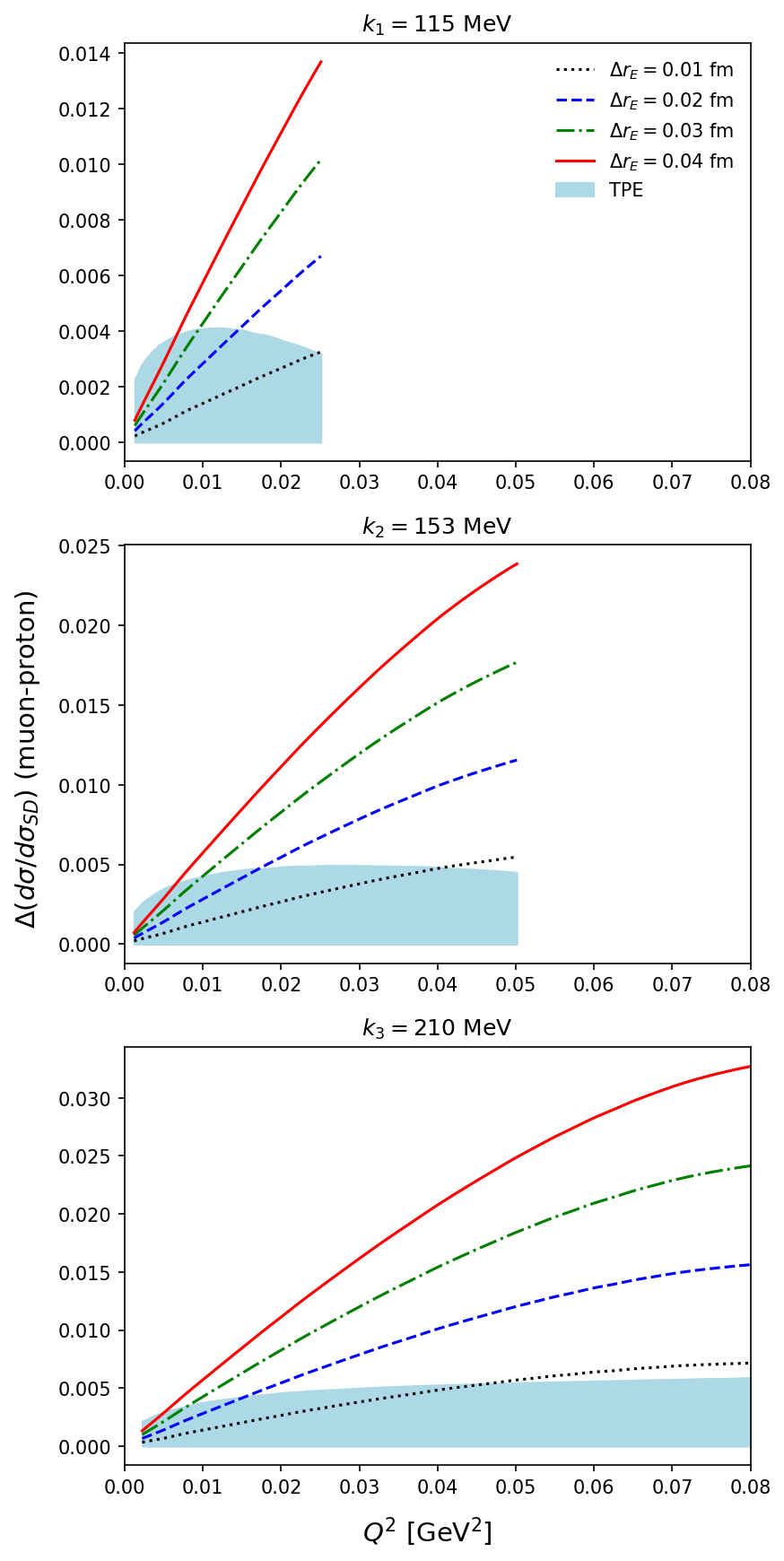} 
\caption{Estimated accuracy of $\mu^- p$ cross section measurement required to discriminate between
different values of the proton radius.
{\it Lines:} Differences between DI$\chi$EFT cross section predictions for proton radii
differing by $\Delta r_E$ (values see legend). {\it Shaded band at bottom:} Size of the TPE contribution
\cite{Tomalak:2015hva}.}
\label{Fig:TPE_effect_in_radius}
\end{figure}
To make this assessment, we use the difference between the cross section predictions for different radii
in Fig.~\ref{Fig:cross_section_radii_mup} as an estimate of the experimental accuracy required
to discriminate between these values of the radii. At each value of $Q^2$ in Fig.~\ref{Fig:cross_section_radii_mup},
we compute the minimal difference between the cross section predictions for radii differing by a 
given $\Delta r_E$, taking the minimum over all pairs of radii with the given $\Delta r_E$,
and taking into account their theoretical uncertainties (i.e., computing the minimal gap between the
theoretical uncertainty bands of the cross section predictions for a given $\Delta r_E$).
The minimal cross section difference computed in this way is independent of the nominal value of $r_E$. 
Figure~\ref{Fig:TPE_effect_in_radius} shows the minimal cross section differences obtained in this way,
for radius differences $\Delta r_E =$ 0.01, 0.02, 0.03 and 0.04 fm, as functions of $Q^2$. 
One observes: (a)~The cross section differences depend strongly on $Q^2$ for fixed $k$.
They depend relatively weakly on $k$ for fixed $Q^2$ (when comparing them at a fixed $Q^2$ that is
kinematically accessible at multiple values of $k$). The main role of $k$ is to define the kinematically
accessible range of $Q^2$. (b)~At low values of $Q^2$, high experimental precision is needed for radius
determination. At $Q^2 \lesssim$ 0.015 GeV$^2$, a relative accuracy $\leq 0.2\%$ is needed for
$\Delta r_E =$ 0.01 fm, independently of $k$. (c)~The demands on the experimental accuracy decrease at higher $Q^2$.
At $Q^2 \sim$ 0.05 GeV$^2$, a relative accuracy $\sim 0.5\%$ is needed for $\Delta r_E =$ 0.01 fm.

Another factor to consider is the uncertainty in the theoretical calculation of the TPE
correction \cite{Tomalak:2015hva}.  Depending on the kinematics, this contribution to the cross
section can be crucial for determining the radius with the necessary precision.
Figure~\ref{Fig:TPE_effect_in_radius} compares the value of the TPE correction with the
predicted cross section differences for a given radius difference (note that the plots show
the estimated total value of the TPE correction, not its uncertainty).  The theoretical uncertainty of the
TPE correction is not well known; however, we can assess how an assumed theoretical uncertainty of
the TPE correction would impact on the overall uncertainty of the radius extraction. For the lowest
beam momentum, $k =$ 115 MeV, the size of the TPE correction is larger than the variation of the
cross section prediction for $\Delta r_E$ = 0.01 fm. The TPE correction thus has a decisive influence
on the radius extraction in this kinematics.  The situation becomes more favorable at higher beam
momenta, where the TPE correction is comparable or smaller than the cross section variation for
$\Delta r_E$ = 0.01 fm.

Overall, our analysis suggests that the optimal kinematics for proton radius determination at MUSE
with the DI$\chi$EFT method is at the highest beam momentum, $k =$ 210 MeV, using momentum transfers
$Q^2 \sim$ 0.05--0.08 GeV$^2$, at the upper end of the experimentally accessible range. In this setting,
the experimental precision required for radius determination with $\Delta r_E$ = 0.01 fm is estimated
at $\approx 0.8\%$. The final uncertainty of the radius extraction depends on the
theoretical uncertainty of the TPE correction, which is not known at present.
%
%
\begin{figure}[t]
\includegraphics[width=0.95\columnwidth]{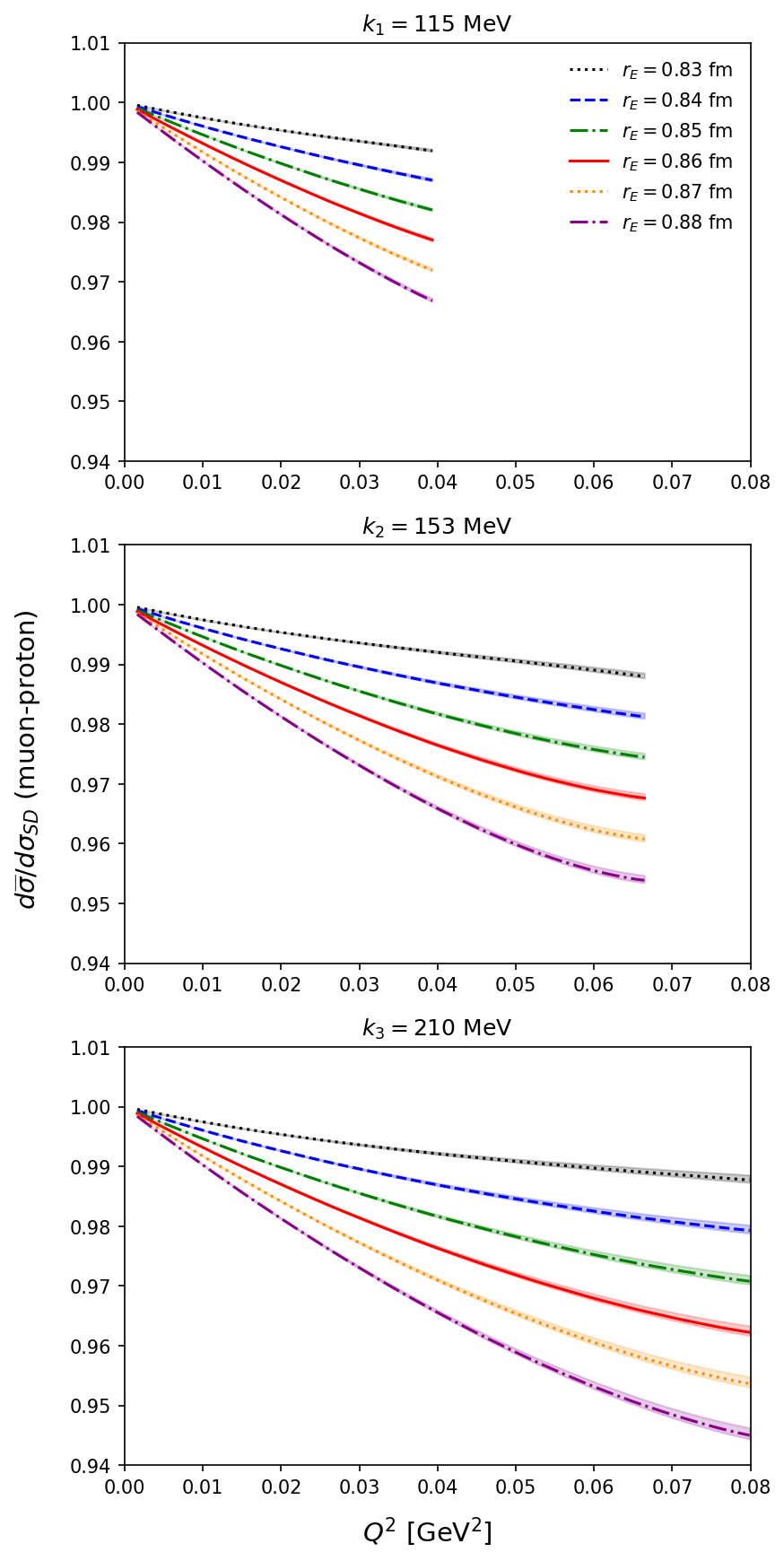} 
\caption{DI$\chi$EFT predictions for the average of $\mu^+$ and $\mu^-$ elastic scattering
cross sections Eq.~(\ref{sigma_charge_average}) in MUSE kinematics, normalized to the standard dipole.
{\it Lines, shaded bands:} Same notation as in Fig.~\ref{Fig:cross_section_radii_mup}.}
\label{fig:cross_section_OPE_mup}
\end{figure}  
An alternative method for proton radius extraction with $\mu p$ scattering uses the average of $\mu^+$
and $\mu^-$ cross sections,
\begin{align}
\bar{\sigma} \equiv [\sigma(\mu^+ p) + \sigma(\mu^- p)]/2,
\label{sigma_charge_average}
\end{align}
in which the TPE correction cancels due to its charge dependence.
The same DI$\chi$EFT analysis of radius sensitivity and optimal kinematics
as above can be performed in this case. The cross section prediction is now given by the one-photon
exchange cross section Eq.~(\ref{Eq:X-sec_1gamma}). Figure~\ref{fig:cross_section_OPE_mup} shows
the predicted cross section and its theoretical uncertainty. The assessment of the optimal $Q^2$ values
is the same as for $\mu^+ p$ above.
\subsection{Cross section prediction for nominal radius}
%
%
\begin{figure}[t]
\includegraphics[width=0.95\linewidth]{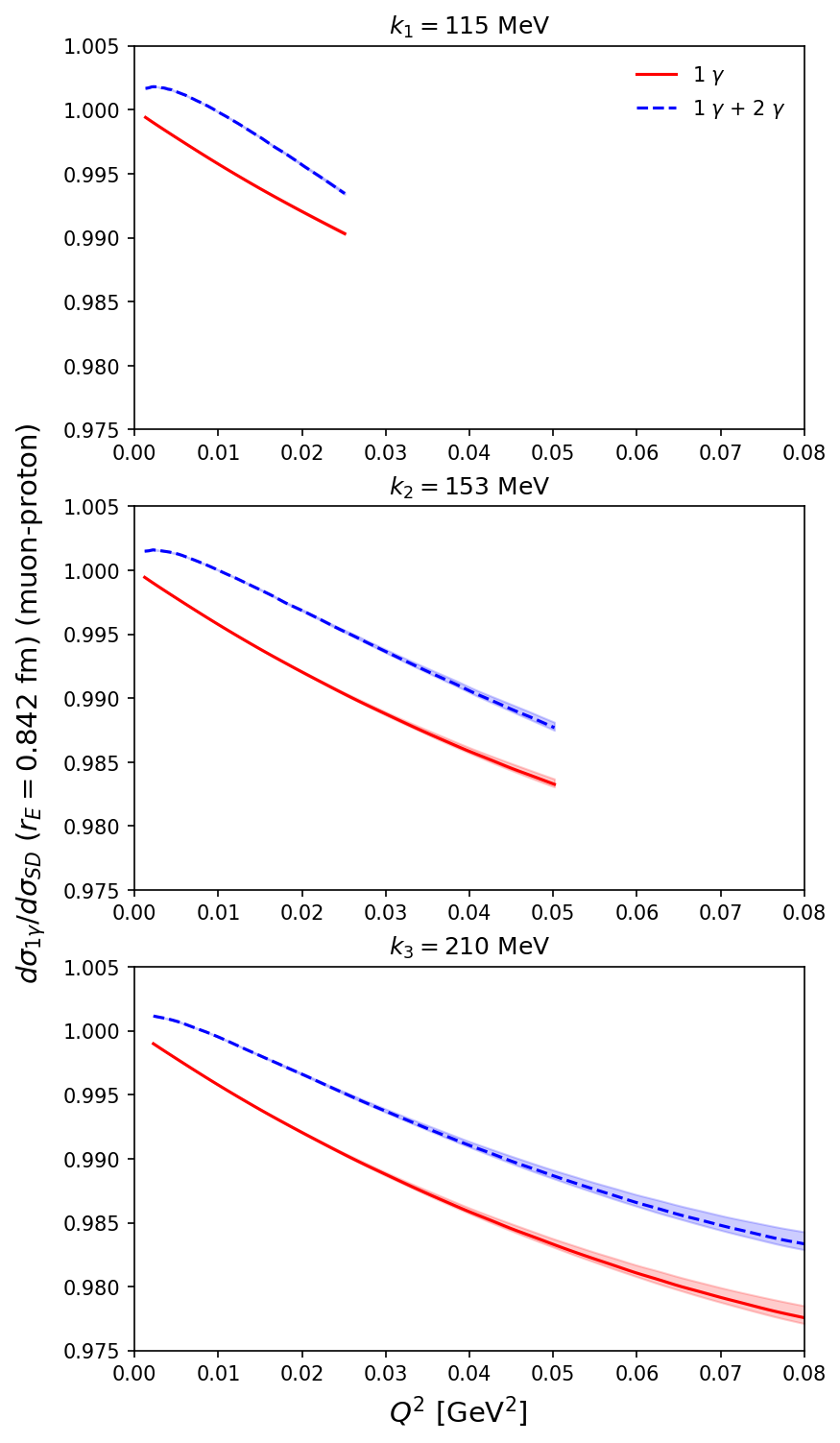} 
\caption{DI$\chi$EFT predictions for $\mu p$ elastic scattering cross at MUSE
for the nominal proton charge radius $r_E = 0.842(2)~\text{fm}$.
{Dashed line and shaded band (blue):} $\mu^- p$ cross section, including
one- and two-photon exchange contributions. 
{Solid line and shaded band (red):} Average of $\mu^+ p$ and $\mu^- p$ cross sections,
given by the one-photon exchange contribution.}
\label{Fig:prediction_MUSE_X-sec}
\end{figure}  
The present study focuses on the prospects for extracting the proton radius from $\mu p$ scattering
experiments at MUSE. The proton radius can also be extracted from atomic spectroscopy and $ep$
scattering experiments. In this context we can use DI$\chi$EFT to predict the $\mu p$
cross section expected for a given value of the radius and its theoretical uncertainty.
For reference, we give here the prediction for the $\mu p$ cross section with the proton charge
radius obtained in the previous DI$\chi$EFT analysis of $ep$ scattering
results \cite{Alarcon:2018zbz,Alarcon:2020kcz}
\begin{align}
r_E = 0.842(2)~\text{fm}.
\end{align}
Figure~\ref{Fig:prediction_MUSE_X-sec} shows the predicted $\mu^- p$ cross section in MUSE kinematics,
Eq.~(\ref{Eq:TP-Correction}), which includes the TPE correction; and the charge-averaged cross
section, Eq.~(\ref{sigma_charge_average}) in which the TPE correction cancels.

\subsection{Comparison of $ep$ and $\mu p$ scattering}
%
%
\begin{figure}[t]
\includegraphics[width=0.9\linewidth]{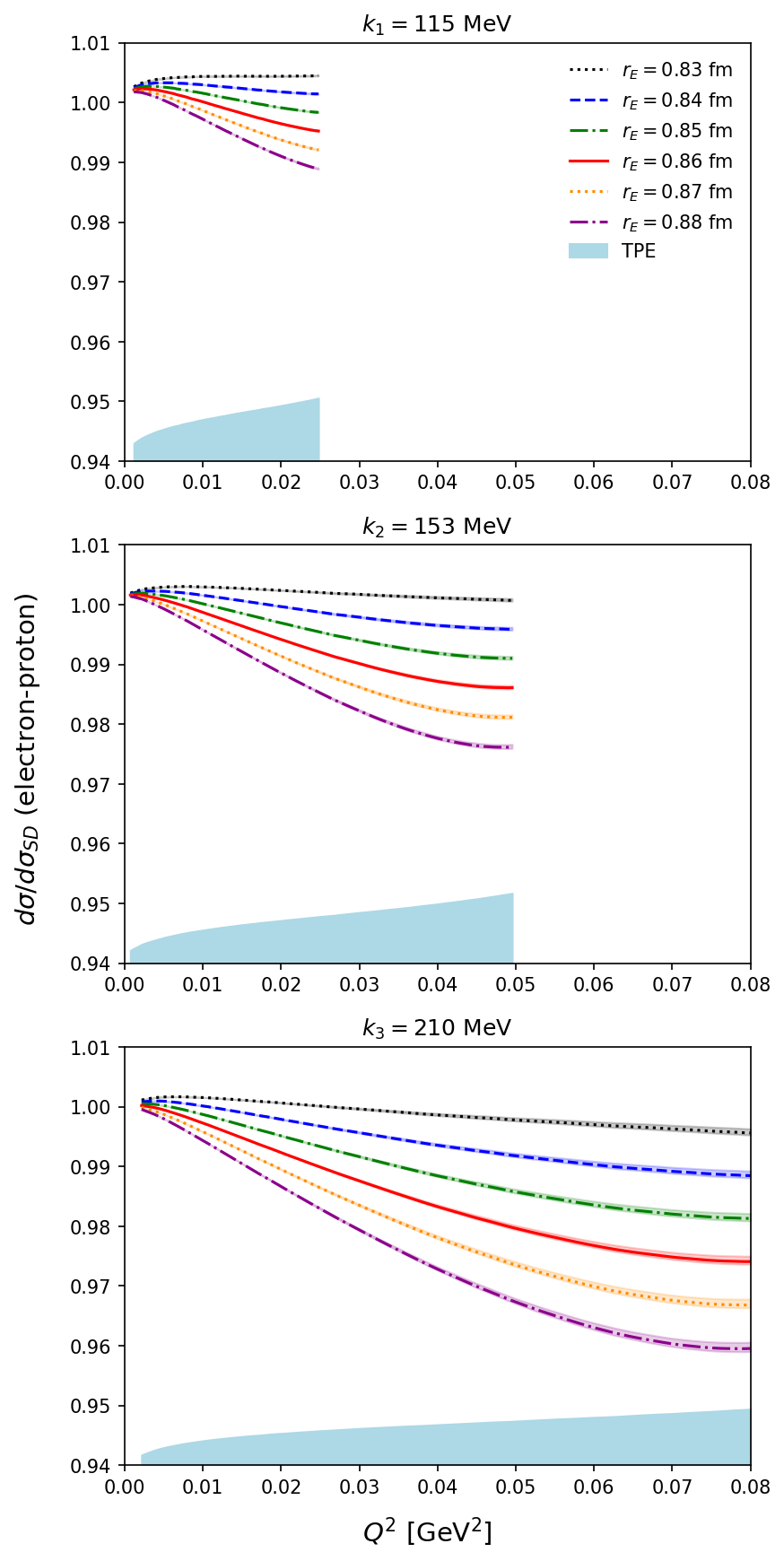} 
\caption{DI$\chi$EFT predictions for the differential cross section of $ep$ scattering for several
assumed value of the proton radius
(compare with Fig.~\ref{Fig:cross_section_radii_mup} for $\mu p$ scattering).
The cross section predictions include the TPE correction and
are normalized by the standard dipole cross section without TPE effects.
{\it Solid lines:} Nominal DI$\chi$EFT predictions for the assumed value of the proton radius. {\it
Bands around solid lines:} Intrinsic theoretical uncertainty of DI$\chi$EFT prediction (unrelated to
assumed proton radius).
{\it Blue band at bottom:} TPE contribution to cross section \cite{Tomalak:2015aoa}.}
\label{fig:ep_cross_section_radii}
\end{figure}
It is interesting to compare the prospects for proton radius extraction in $ep$ and $\mu p$ scattering
in the same kinematics. The MUSE experiment will measure both $ep$ and $\mu p$ scattering, and methods
for proton radius extraction were studied intensively in earlier $ep$ scattering experiments.
Characteristic differences between $ep$ and $\mu p$ occur in the TPE
effects \cite{Tomalak:2015hva,Tomalak:2015aoa} and in the role of the
magnetic FF. We exhibit them by repeating the DI$\chi$EFT analysis for $ep$ scattering and comparing
with the $\mu p$ results.

Figure~\ref{fig:ep_cross_section_radii} shows the DI$\chi$EFT predictions for the $ep$ cross section
for a range of assumed values of the proton radius, in the same style as Fig.~\ref{Fig:cross_section_radii_mup}
for $\mu p$. One observes: (a) The TPE corrections have different kinematic dependence
in $ep$ than in $\mu p$ scattering \cite{Tomalak:2015hva,Tomalak:2015aoa}. In $ep$ they increase strongly with
$Q^2$ at fixed $k$, and decrease with $k$ at fixed $Q^2$ In $\mu p$ the dependencies are much weaker.
(b)~The size of the TPE corrections relative to the variation of the cross section with the radius is
much larger in $ep$ than in $\mu p$, especially at low beam momenta. At $k =$ 115 MeV and $Q^2 = 0.02$ GeV$^2$,
the size of TPE correction amounts to a change of the radius $\Delta r_E \approx$ 0.03 fm in $ep$ scattering,
compared to $\Delta r_E \approx$ 0.015 fm in $\mu p$ scattering in the same kinematics.
(c)~Overall, the different size and kinematic dependence of the TPE corrections causes a different
$Q^2$-dependence of the cross section for $ep$ and $\mu p$ scattering at low $Q^2$.

%
%
\begin{figure}[t]
\includegraphics[width=0.95\columnwidth]{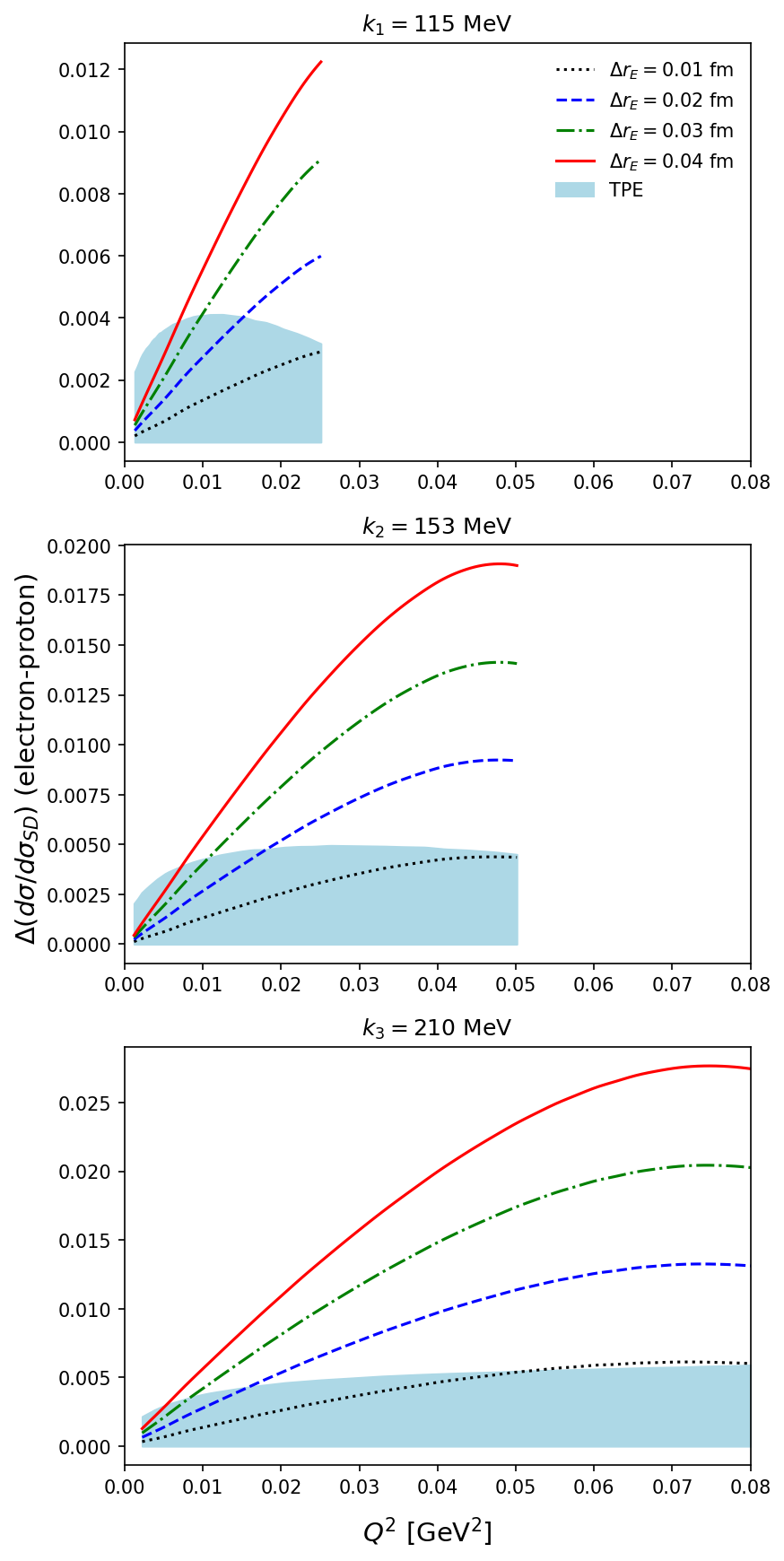} 
\caption{Estimated accuracy of $e^- p$ cross section measurements required to discriminate between
different values of the proton radius
(compare with Fig.~\ref{Fig:TPE_effect_in_radius} for $\mu p$ scattering).
{\it Lines:} Differences between DI$\chi$EFT cross section predictions for proton radii
differing by $\Delta r_E$ (values see legend). {\it Shaded band at bottom:}
Size of the TPE contribution \cite{Tomalak:2015aoa}.}
\label{fig:required_accuracy_ep}
\end{figure}
Figure~\ref{fig:required_accuracy_ep} shows the estimated accuracy of the $e^-p$ cross section measurement
required for discriminating between different values of the radius, in the same style as
Fig.~\ref{Fig:TPE_effect_in_radius} for $\mu^- p$. The optimal kinematics for radius extraction
in $ep$ scattering at MUSE can be determined in the same way as for $\mu p$.
The results of Figure~\ref{fig:required_accuracy_ep} show that the experimental accuracy required
for radius extraction from $ep$ scattering is least at the highest beam momentum $k = 210$ MeV.
The optimal $Q^2$ value determined by the trade-off between radius sensitivity and theoretical uncertainty is
at $Q^2 \sim $ 0.065 GeV$^2$, slightly below the kinematic limit.

%
%
\begin{figure}[t]
\includegraphics[width=0.9\linewidth]{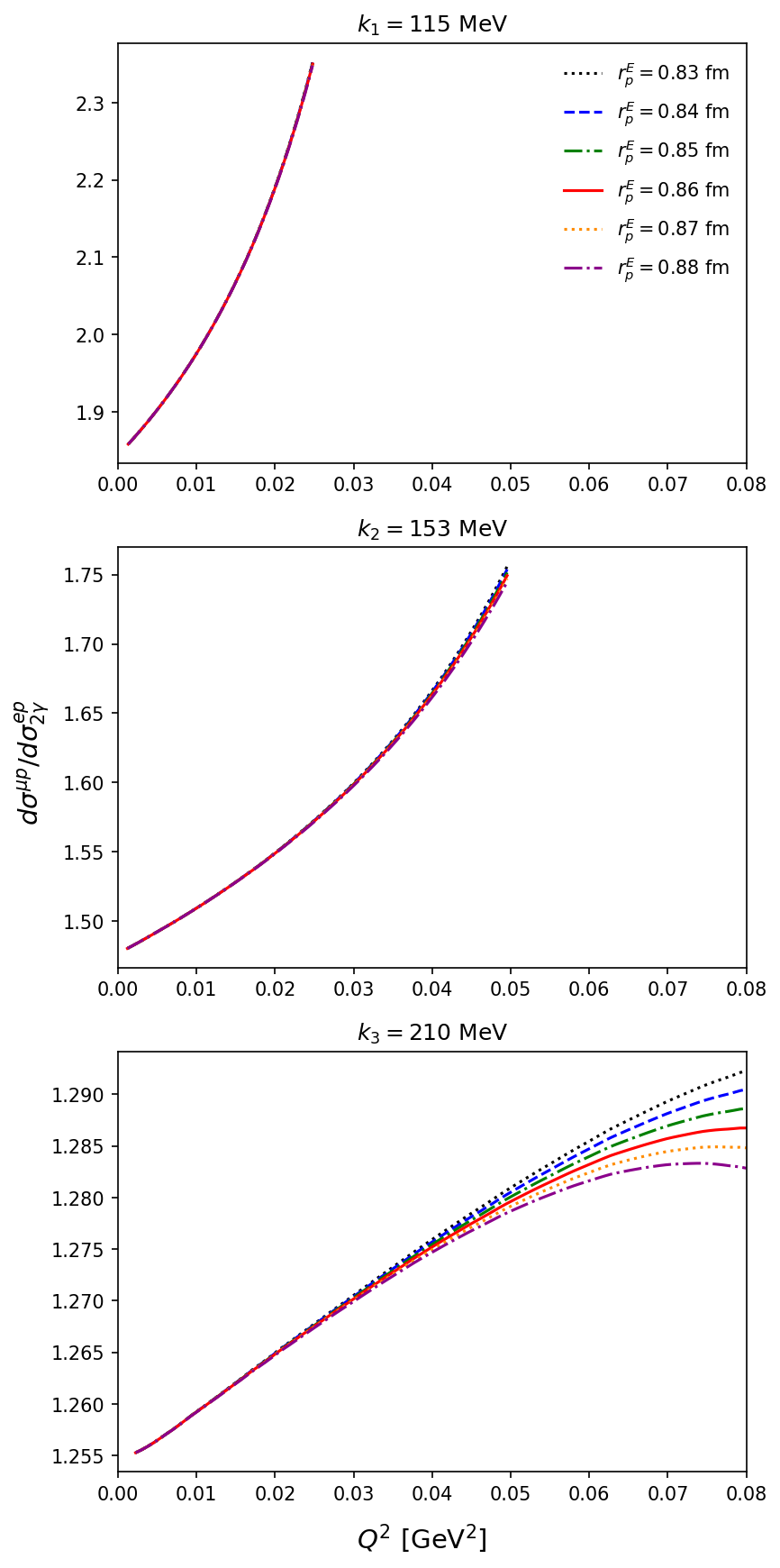}
\caption{DI$\chi$EFT predictions of the ratio of $\mu p$ to $ep$ elastic scattering cross sections
$\sigma$, Eq.~(\ref{Eq:TP-Correction}), including the TPE corrections,
for several assumed values of the proton radius.}
\label{Fig:ratio_mup_ep}
\end{figure}  
Figure~\ref{Fig:ratio_mup_ep} shows the ratio of the $\mu p$ to $ep$ cross sections (including the TPE
corrections) for the same set of assumed proton radii. The ratio directly expresses the different
$Q^2$-dependence of the $ep$ and $\mu p$ cross sections. Its magnitude and $Q^2$-dependence are determined
by the kinematic factors in the one-photon-exchange cross section Eq.~(\ref{Eq:X-sec_1gamma}) et seq.
One observes that the ratio is remarkably insensitive to the proton charge radius in this kinematic regime,
especially at the lower values of $k$. Only at $k =$ 210 MeV the differences between the radii become
visible at the largest $Q^2$ values.
 
The results of Fig.~\ref{fig:ep_cross_section_radii} show that the TPE corrections play a much larger
role in proton radius extraction from $ep$ scattering than $\mu p$ scattering, and that they limit the theoretical
uncertainty of the extracted radius. With the DI$\chi$EFT method, the influence of TPE corrections in
$ep$ can be minimized by using the data at the highest beam momentum $k =$ 210 MeV and momentum transfers
in the range $Q^2 \sim$ 0.03--0.08 GeV$^2$ for radius extraction. In this kinematics the cross section
shows good sensitivity to the proton charge radius, the theoretical uncertainty of the DI$\chi$EFT predictions
is small, and the size of the TPE corrections amounts to a shift of the radius $\Delta r_E \sim$ 0.01 fm
(see Fig.~\ref{fig:ep_cross_section_radii}). The ability to recruit higher-$Q^2$ data for radius extraction
with DI$\chi$EFT is thus even more advantageous in $ep$ than in $\mu p$ scattering.

%
%
\begin{figure}[t]
\includegraphics[width=0.8\linewidth]{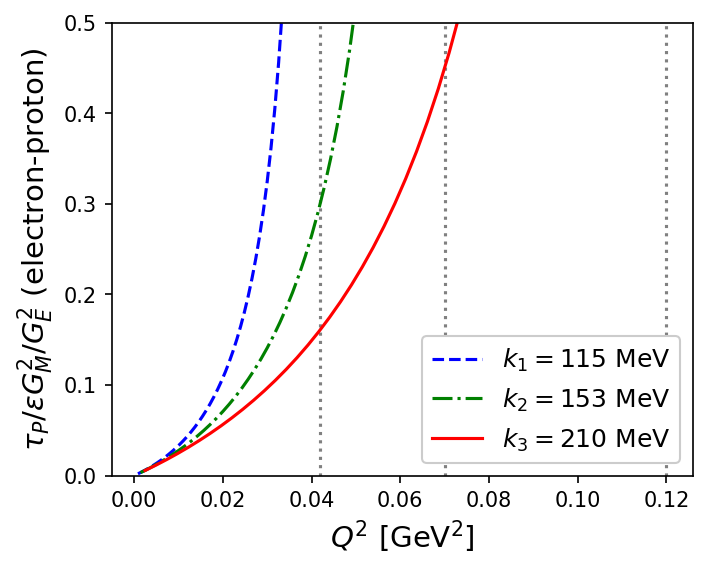} 
\caption{Ratio of magnetic and electric contributions to the $ep$ elastic scattering cross section,
$(\tau_P / \epsilon) G_M^2/G_E^2$, in MUSE kinematics
(compare with Fig.~\ref{Fig:GM_GE_ratio_mup} for $\mu p$ scattering).
The vertical dotted lines represent the kinematic upper limits of $Q^2$ at the
given beam momentum $k$, Eq.~(\ref{Q2_limits}).}
\label{Fig:GM_GE_ratio_ep}
\end{figure}
An important difference between $ep$ and $\mu p$ scattering appears in the contribution of the
magnetic FF at large momentum transfers, at the upper end of the allowed kinematic range.
In $ep$ scattering $\epsilon = 0$ at $Q^2 \sim Q^2_{\rm max}$ (if one neglects the electron mass),
while in $\mu p$ scattering $\epsilon$ attains a finite value, see Eq.~(\ref{epsilon_Q2max}).
In $ep$ scattering the one-photon exchange cross section for $Q^2 \rightarrow Q^2_{\rm max}$
is therefore dominated by $G_M$. Figure~\ref{Fig:GM_GE_ratio_ep} shows the ratio of magnetic
and electric contributions to the one-photon-exchange cross section for $ep$ scattering,
in the same way as Fig.~\ref{Fig:GM_GE_ratio_mup} for $\mu p$. One sees that the magnetic
contribution to the cross section is substantially larger in $ep$ than $\mu p$ already
for $Q^2$ in the middle of the kinematic range. This circumstance must be taken into account when
assessing the sensitivity of the cross section to $r_E$ in DI$\chi$EFT, and one should remain
in the region where the cross section is not dominated by $G_M$.
We have quantified the impact of the uncertainty in $G_M$ on the proton radius extraction
from $ep$ scattering in the same way as for $\mu p$ (see Sec.~\ref{subsec:sensitivity}),
using the DI$\chi$EFT framework and the experimental uncertainty of $G_M$ obtained in an
earlier analysis \cite{Alarcon:2020kcz}. We find that the current experimental uncertainty of $G_M$
produces a relative variation of the cross section of at most $\sim 0.05\%$ in the range covered
by Fig.~\ref{fig:ep_cross_section_radii}. The uncertainty from $G_M$ is thus not a limiting
factor of the DI$\chi$EFT-based radius extraction from $ep$ scattering data in MUSE kinematics.

%
%
\begin{figure}[t]
\includegraphics[width=0.9\linewidth]{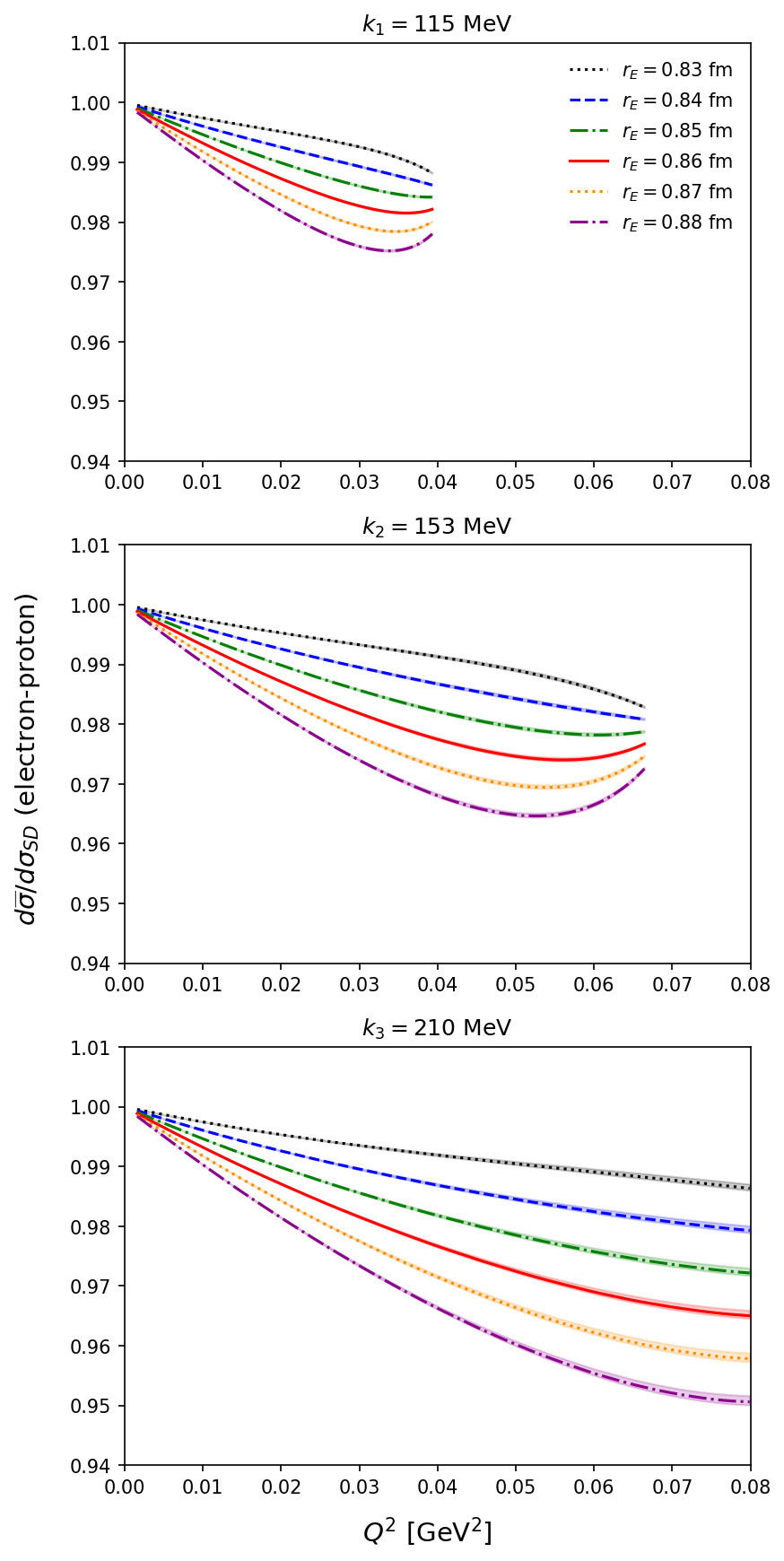} 
\caption{DI$\chi$EFT predictions for the average of $e^+$ and $e^-$ elastic scattering
cross sections, as in Eq.~(\ref{sigma_charge_average}) for $\mu^+ p$ and $\mu^- p$,
in MUSE kinematics
(compare with Fig.~\ref{fig:cross_section_OPE_mup} for $\mu p$ scattering).}
\label{fig:ep_cross_section_average}
\end{figure}
Because of the large TPE corrections in $ep$ scattering, it would be an attractive option to perform
the proton radius extraction with the average of $e^- p$ and $e^+ p$ cross sections
[see Eq.~(\ref{sigma_charge_average}) for $\mu^\pm p$], in which the TPE effects cancel.
In this case the cross section is accurately described by the one-photon-exchange formula,
and the analysis greatly simplifies. Figure~\ref{fig:ep_cross_section_average} shows the
DI$\chi$EFT predictions for the charge radius dependence of the charge-averages cross section
$\bar{\sigma}$ in $e^\pm p$ scattering in MUSE kinematics.

\section{Conclusions}
In this work we have used the DI$\chi$EFT framework to study the prospects for the proton radius
extraction from $\mu p$ scattering at MUSE. The principal conclusions are:

(i) When extracting the radius from fits to cross section data at low momentum transfers $Q^2 <$ 0.01 GeV$^2$,
the TPE corrections need to be included with high precision. At $Q^2 <$ 0.01 GeV$^2$, the estimated absolute
size of TPE correction amounts to a shift of the extracted radius by 0.03--0.04 fm. Any theoretical uncertainty
of the TPE correction will influence the extracted radius proportionally.

(ii) The DI$\chi$EFT method allows one to extract the radius from fits to cross section data at higher
momentum transfers $Q^2 \sim$ few times 0.01 GeV$^2$ in the MUSE kinematic range. This is advantageous
experimentally, because the higher sensitivity of the cross section to the radius lowers the demands
on the experimental precision of the cross section measurement. It is also advantageous theoretically,
as it reduces the influence of the TPE correction on the radius extraction.

(iii) The optimal kinematics for the DI$\chi$EFT-based radius extraction at MUSE is $k = 210$ MeV
and $Q^2 \sim$ 0.05--0.08 GeV$^2$, at the upper end of the kinematic coverage. It is determined by the trade-off
between theoretical effects -- the sensitivity of the cross section to the radius, the uncertainty
of the DI$\chi$EFT FF predictions, and the TPE correction. In this kinematics, even a 100\% uncertainty of
the TPE correction would shift the extracted radius only by 0.01 fm. An experimental precision of $\leq$ 0.5\%
is required for determining the radius with 0.01 fm accuracy. Such accuracy would be sufficient for solving
the proton radius puzzle.

(iv) In $ep$ scattering in MUSE kinematics, the TPE corrections are generally larger that in $\mu p$,
and the advantages of using the DI$\chi$EFT method with higher-$Q^2$ data for radius extraction are
even more compelling. The ratio of same-charge $ep$ and $\mu p$ cross sections is predicted to be
practically independent of the proton radius and can be used for validation of the analysis.

Our findings affirm the need for accurate theoretical estimates of the TPE corrections in elastic
$\mu p$ and $e p$ scattering. If the radius extraction is performed using the DI$\chi$EFT framework
and data at finite momentum transfers $Q^2 \sim$ 0.05--0.08 GeV$^2$, as recommended here, efforts
should focus on improving the TPE estimates in this kinematic region. At these finite values of $Q^2$
the constraints on the TPE amplitude from the limit of forward scattering ($Q^2 = 0$) are less
restrictive, and the calculations become more dependent on dynamical
assumptions \cite{Tomalak:2015hva,Tomalak:2015aoa}.
Methods based on the $1/N_c$ expansion of QCD could enable systematic calculations of TPE effects with
controlled theoretical uncertainties; see Refs.~\cite{Goity:2022yro,Goity:2023sph} for recent developments.

The DI$\chi$EFT framework used in the present study is a general method that could be improved through
further development. In particular, the theoretical uncertainties of the FF predictions could be reduced
by using a more flexible parametrization of the high-mass states in the spectral function,
Eq.~(\ref{spectral_pipi_highmass}), and constraining it with spacelike nucleon FF data.
In the analysis here we have used the version of Ref.~\cite{Alarcon:2018irp}, where the high-mass states
are described by a single effective pole, which permits simple uncertainty estimates and is
sufficient for the MUSE $Q^2$-range. A version with a more elaborate description of the high-mass states,
using multiple poles with randomized parameters for the uncertainty estimates, was described
in Ref.~\cite{Alarcon:2022adi} and could be employed for elastic scattering analysis and
radius extraction at higher $Q^2 \lesssim$ 1 GeV$^2$.
\section*{Acknowledgments}
We thank A.~Gramolin for making available a computer code that was used to validate the numerical calculation
of the one-photon exchange cross section, and P.~Blunden and W.~Melnitchouk for helpful correspondence about
two-photon exchange effects.

J.M.A.\ acknowledges support from the Spanish MICINN grant PID2019-106080GB-C21.
This material is based upon work supported by the U.S.~Department of Energy, 
Office of Science, Office of Nuclear Physics under contract DE-AC05-06OR23177.
\bibliography{muse}
\end{document}